\documentclass[twocolumn,prd,showpacs,10pt]{revtex4}
\usepackage{graphicx}
\usepackage{dcolumn}
\usepackage{bm}

\newcommand{\square}{\kern1pt\vbox{\hrule height 1.2pt\hbox{\vrule
width 1.2pt\hskip 3pt
\vbox{\vskip 6pt}\hskip 3pt\vrule width 0.6pt}\hrule
height 0.6pt}\kern1pt}
\newcommand{\beq}{\begin{equation}}
\newcommand{\beqn}{\begin{eqnarray}}
\newcommand{\eeq}{\end{equation}}
\newcommand{\eeqn}{\end{eqnarray}}

\begin{document}

\title{Cosmological evolution of general scalar fields in a brane-world cosmology}\author{Shuntaro Mizuno$^1$, Seung-Joo Lee$^2$,
and Edmund. J.~Copeland$^2$}
\address{$^1$Department of Physics, Waseda University, Okubo 3-4-1,
Shinjuku, Tokyo 169-8555, Japan}
\address{$^2$ Department of Physics and Astronomy, University of Sussex,
Falmer, Brighton BN1 9QJ, United Kingdom}
\date{\today}

\begin{abstract}
We study the cosmology of a general scalar field and barotropic fluid during the early stage of a 
brane-world where the Friedmann constraint is dominated by the square of the energy density.
Assuming both the scalar field and fluid are confined to the brane, we find a range of behaviour depending on the form of the potential. Generalising an approach developed for a standard Friedmann cosmology, in \cite{delaMacorra:1999ff}, we show that the potential dependence $V(\phi)$ can be  described through a parameter 
$\lambda \equiv -\sqrt{2} m_5^{3/2} V'/(\sqrt{H}V)$, where $m_5$ is the 5-dimensional Planck mass,  $H$ is the Hubble parameter and $V' \equiv \frac{dV}{d\phi}$. For the case where $\lambda$ asymptotes to zero, we show that the solution exhibits stable inflationary behaviour. On the other hand if it approaches a finite constant, then $V(\phi) \propto \frac{1}{\phi^2}$. For $\lambda \to \infty$ asymptotically, we find examples where it does so both with and without oscillating. In the latter case, 
the barotropic fluid dominates the scalar filed asymptotically.
Finally we point out an interesting duality which leads to identical evolution equations in the high energy $\rho^2$ dominated regime and the low energy $\rho$ dominated regime.

\end{abstract}

\vskip 1pc \pacs{pacs: 98.80.Cq}
\maketitle

%

\section{Introduction}
Scalar fields play very important roles in cosmology,
especially in the early stage of the universe.
The Inflationary scenario \cite{inflation} relies on the potential energy of a scalar `inflaton' field to drive a period of early universe acceleration. It is thought that the early universe could well have also been characterised by a series of phase transitions, in which topological defects could have been formed \cite{Cosmic_string,cosmicstrings}. Moreover in context of the string theory, 
the natural values of the 
gauge and gravitational couplings in our 4d universe
are explained by the dynamics of `moduli' scalar fields. 
\cite{Green:sp,moduli_stabilization}. More recently, potentials of light scalar fields have been invoked as a possible source of the Dark Energy which appears to be dominating the dynamics of our Universe today. 
The success of this approach has often been established by demonstrating the existence of attractor solutions by investigating  the asymptotic dynamics  
of the system being studied. For example in the context of  usual 4D Einstein gravity, particular potentials have been discussed  
\cite{Ratra_Peebles,Steinhardt,CLW,trac,ferreira,Liddle,sahni}, whereas 
others  have adopted an approach which does not specify a particular potential  \cite{delaMacorra:1999ff,Ng:2001hs,Corasaniti:2002vg}. 
However, recent higher-dimensional unification scenarios 
based on  brane-worlds suggest that the gravitational law
could be different from Einstein's during the early stages 
of our Universe. In the brane-world scenario, our Universe is 
a four dimensional hypersurface embedded in higher dimensions.
Standard-model particles are confined to the brane, while gravity
can propagate in the higher-dimensional bulk space
\cite{old_brane,H_W,Lukas:1997fg,Lukas:1998yy,Lukas:1998tt,A_D_D,R_S}.

Both the models based on Horava-Witten and Randall-Sundrum II (R-S II) are interesting because they lead to a new type of compactification with gravity.
They also make concrete predictions of how conventional gravity will be modified at high-energy scales. Many authors have discussed the cosmology associated with these scenarios
\cite{Lukas:1998dc,Lukas:1999yn,Copeland:2001zp,Maartens,Langlois,Brax,b_review}.
When we consider the homogeneous and isotropic cosmology based on these
brane world models, the difference from  conventional cosmology can be related to the appearance of
two new terms in the Friedmann equation, i.e., the quadratic term
of the energy-momentum and a dark radiation term \cite{SMS,Binetruy,Mukohyama}.
The allowed impact of dark radiation, is constrained
by nucleosynthesis \cite{econstraint}, with the most important change in
the scalar field dynamics being due to the appearance of the quadratic
energy density term.

In this paper, we study the dynamics of 
a scalar field which is confined to the brane
in the context of a R-S II brane-world. 
As well as the scalar field, we include a barotropic fluid on the brane (which could be radiation or matter).
Since the modification of the cosmic expansion law
arising through the quadratic energy density term becomes 
important during the early stage of the universe, we concentrate on this stage.
In order to classify the asymptotic behaviour of the solutions, we follow 
the model independent approach as first proposed in 
\cite{delaMacorra:1999ff,Ng:2001hs}. Earlier work investigating the impact of these $\rho^2$ corrections has concentrated on models with particular potentials \cite{Copeland:2000hn,Sahni:2001qp,Majumdar:2001mm,Nunes:2002wz,Lidsey:2003sj,Tsujikawa:2003zd,Sami:2004ic}.  We intend to go beyond that in this paper, and in passing note that the analysis complements our previous discussion on scaling solutions in $\rho^2$ cosmologies \cite{MMY}.

The organization of the paper is as follows.
In Sec.~II, we review the key equations of motion and 
define variables which will allow us 
to analyse the asymptotic behaviour of this system and point out an intriguing duality with the equations obtained in the context of scaling solutions in standard Friedman cosmologies \cite{CLW,delaMacorra:1999ff}.  In Sec.~III we obtain the class of attractor solutions which exist for constant $\lambda$. We then extend the analysis to the cases where $\lambda$ evolves and find classes of solutions corresponding to inflationary regimes (Sec.~IV);
scalar field kinetic energy dominated regimes (Sec.~V), and  
oscillating regimes (Sec.~VI).
We summarise our results in Section VII.

\section{Equations of motion}
We analyze the dynamics of a scalar field in a Randall-Sundrum II
(R-S II) brane-world \cite{R_S}, because the model is simple and concrete. 
However, we expect  our key results to also hold in other  brane-world models, in which a quadratic term
in the energy-momentum tensor generically appears. 
In the R-S II model, even though the
extra-dimension is not compactified, gravity is confined  to
the brane at low energy, resulting in Newtonian gravity in our world which is described by the
intrinsic metric of the 4-dimensional  brane spacetime. By use of Israel's
junction condition and assuming
$Z_2$  symmetry, the gravitational equations on the 3-brane
are given by\cite{SMS}
\beqn
^{(4)}G_{\mu\nu}&=&-^{(4)}\Lambda g_{\mu\nu}+\kappa_4^2
T_{\mu\nu}+\kappa_5^4 \Pi_{\mu\nu}-E_{\mu\nu},
\label{geq}\nonumber\\
\Pi_{\mu\nu} &\equiv& -\frac{1}{4}T_{\mu\alpha}T_{\nu}^{\alpha}+
\frac{1}{12} T T_{\mu\nu} + \frac{1}{8}g_{\mu\nu} T_{\alpha \beta}
T^{\alpha \beta}-\frac{1}{24} g_{\mu\nu} T^2,
\nonumber\\
\eeqn
where $^{(4)}G_{\mu\nu}$ is the Einstein tensor with respect
to the intrinsic metric $g_{\mu\nu}$, $^{(4)}\Lambda$ is the 
4-dimensional cosmological constant, $T_{\mu\nu}$
represents the energy-momentum tensor of matter fields confined to the
brane  and
$\Pi_{\mu\nu}$ is the local correction term
as a result of a brane embedded in the bulk. 
$E_{\mu\nu}$ is a part
of the 5-dimensional Weyl tensor and carries information about the bulk
geometry.
$\kappa_4^{\;2}=8\pi G_4$ and
$\kappa_5^{\;2}=8\pi G_5$ are 4-dimensional and 5-dimensional
gravitational constants,  respectively. In what follows, we use the
4-dimensional Planck mass $m_4 \equiv \kappa_4^{~-1}
=(2.4\times10^{18}{\rm GeV})$ and the 5-dimensional
Planck mass $m_5\equiv
\kappa_5^{~-2/3}$.
$m_5$ is related to $m_4$ in terms of the brane tension $\sigma$
as $ \sigma m_4^2 = 6 m_5^6$, and could be much smaller than $m_4$.

Assuming the Friedmann-Robertson-Walker
spacetime in our brane world,  we find the effective Friedmann equations
arising from Eq.(\ref{geq}) are
\beqn
H^2+{k\over
a^2} &=&\frac{1}{3}{}^{(4)}\Lambda +\frac{1}{3m_4^{\;2}}
\rho+\frac{1}{36m_5^{\;6}} \rho^2+\frac{{\cal C}}{a^4}
\label{fr1}
\\
\dot{H}-{k\over a^2}&=&-\frac{1}{2m_4^{\;2}} \left(P+\rho\right)
-\frac{1}{12m_5^{\;6}} \rho\left(P+\rho\right)-\frac{2{\cal C}}{a^4}
\label{fr2}
\eeqn
where $a$ is a scale factor of the universe, $H=\dot{a}/a$ is its Hubble
parameter, $k$ is a curvature constant,
$P$ and
$\rho$ are the total pressure and  energy density of matter fields,
respectively.
${\cal C}$ is a constant related to ``dark radiation"
coming from $E_{\mu\nu}$\cite{Binetruy,Mukohyama}.
In what follows, we consider only a 
flat Friedmann model ($k=0$). We also assume that $^{(4)}\Lambda $
vanishes at least in the early stage of the universe
and ${\cal C}$ is negligible since it should be diluted by inflation
and is strongly constrained at the time of Nucleosynthesis \cite{econstraint}. 
(However, it must be noted that if we consider perturbations around the
back ground solutions in this paper, we must consider the term arising from
$E_{\mu\nu}$.)

For matter fields on the brane, we consider  a scalar field $\phi$
with a potential $V(\phi)$
as well as a barotropic fluid with an equation of state
$P_B = (\gamma -1) \rho_B$, where $\gamma$ is 
an adiabatic index.
(We will mainly be  considering the case of radiation with 
$\gamma = 4/3$, since we are interested in the early stage of the universe
before nucleosynthesis.) 
We will also assume that the two fluids do not couple to each other explicitly.

Although a 5-dimensional scalar field living in the
bulk\cite{MW} may also appear in other brane-world scenarios, 
in this paper we only consider  a 4-dimensional scalar field 
confined to the brane.
Such a scalar field might originate in the 
condensation of fermionic matter fields  confined to the brane.

As the universe expands, the energy density decreases.
This means that the quadratic term could have been very important in the early stages
of the universe. Comparing the energy density
terms in Eq.~(\ref{fr1}), we find that the quadratic term dominates
when
\beq
\rho > \rho_c \equiv 12 m_5^{~6}/m_4^{~2} .
\eeq

In this regime 
the expansion law of the Universe is clearly modified.
For example, the expansion law in the radiation-dominant era becomes 
$a\propto t^{1/4}$ compared to  the conventional case, 
$a\propto t^{1/2}$. If such a period of $\rho^2$ domination did exist, then
Nucleosynthesis would have provided us with a constraint on $m_5$, since
it must have occured during  the conventional-radiation dominated era in order to explain the
abundances of the light elements. For example, assuming that  the energy density at
$a=a_c$ is  dominated by radiation as $\rho_c \sim   (\pi^2/30) g T_c^4$,
where $g$ is the number of degrees of freedom of light particles,
the temperature of the universe
$T_c$ must be higher than that of nucleosynthesis, i.e. $T_c> T_{NS} \sim
1  \rm{MeV}$. This constraint implies that
$m_5 >1.6 \times 10^4(g/100)^{1/6}
(T_{\rm NS}/1\:\:{\rm MeV})^{2/3} \:\:{\rm GeV}$.
(For the brane tension $\sigma$, this constraint is 
$\sigma^{1/4} > 2.0 {\rm MeV}$).

In the rest of the paper we will be concentrating on the evolution of the universe in the regime dominated by the $\rho^2$ contribution to the energy density. We will find there is a way of performing this analysis in an analogous manner to that already developed for the usual cosmologies  dominated by the linear $\rho$ term \cite{CLW,delaMacorra:1999ff,Ng:2001hs}. In fact we will introduce a set of variables which will lead to exactly the same form of evolution equations for the two regimes, allowing us to relate the attractor solutions that exist in the two regimes through duality transformations.  Of course, in a realistic cosmology, the $\rho^2$ dominated regime is eventually replaced by the linear regime. Any asymptotic solution in the $\rho^2$ regime should be reached before the linear regime takes over at $\rho \sim \rho_c$, and we confirm that  is the case in this analysis. 

In the regime where $\rho \gg \rho_c$, Eq.~(\ref{fr1}) along with conservation of energy and momentum for each fluid component on the brane gives
\beqn
\dot{H} &=& -\frac{1}{12m_5^6} 
(\rho_B + \frac{1}{2}\dot{\phi}^2 + V)(\gamma \rho_B + \dot{\phi}^2),\\
\dot{\rho_B} &=& - 3\gamma H \rho_B,\\
\label{KK}
\ddot{\phi} &=& -3H \dot{\phi} -\frac{dV}{d\phi},
\eeqn
which are subject to the Friedmann constraint given in Eq.~(\ref{fr1})
\beq
\label{quadratic_fri}
H = \frac{1}{6m_5^3} (\rho_B + \frac{1}{2} \dot{\phi}^2 +V),
\eeq
where dots denote derivatives with respect to time.
The energy density and pressure of the homogeneous scalar field
are given by $\rho_\phi=\dot{\phi}^2/2 + V$ and 
$P_\phi=\dot{\phi}^2/2 - V$, respectively.
The effective adiabatic index of the scalar field at any
time,
\beq
\gamma_\phi \equiv \frac{\rho_\phi + P_\phi}{\rho_\phi}
= \frac{\dot{\phi}^2}{\dot{\phi}^2/2 + V},
\eeq
is, in general, time or scale dependent.

In order to analyze the stability of the solutions, 
we modify the variables, $X_{\rm CLW},Y_{\rm CLW}$ first introduced in Ref.~\cite{CLW} to investigate the evolution of a scalar field in a conventional Friedmann cosmology. Defining 
\beqn
\label{def_x}
X &=& \frac{m_5^{-3/2}}{2\sqrt{3}} \frac{\dot{\phi}}{\sqrt{H}},\\
\label{def_y}
Y &=& \frac{m_5^{-3/2}}{\sqrt{6}} \frac{\sqrt{V}}{\sqrt{H}},
\eeqn
where a prime denotes a derivative with respect to 
the logarithm of a scale factor $a$, $N \equiv \ln a$, we see that these are related to those defined in Ref.~\cite{CLW} by 
\beqn
\label{def_xyclw}
\frac{X}{X_{\rm CLW}} &=& \frac{Y}{Y_{\rm CLW}} = \frac{\kappa_5}{\kappa_4} \sqrt{\frac{H}{2}}\\
\eeqn

In terms of these new variables, the equations of motion
read,
\beqn
\label{b_eq_x}
X' &=& -3X + \lambda \sqrt{\frac{3}{2}} Y^2 
+ \frac{3}{2} X [2 X^2 + \gamma (1- X^2 - Y^2)],\nonumber\\
\\
\label{b_eq_y}
Y' &=& -\lambda \sqrt{\frac{3}{2}} XY 
+ \frac{3}{2}Y[2X^2 + \gamma (1- X^2 - Y^2)],\nonumber\\
\\
\label{b_eq_l}
\lambda' &=& -\sqrt{6} \lambda^2 (\Gamma -1) X
+ \frac{3}{2} \lambda [2X^2 + \gamma (1-X^2-Y^2)],\nonumber\\
\eeqn
where generalising the expressions in Refs.~\cite{delaMacorra:1999ff} and~\cite{Steinhardt},
we have defined 
\beqn
\label{def_lambda}
\lambda &\equiv& - \sqrt{2}\frac{m_5^{3/2}}{\sqrt{H}} \frac{dV/d\phi}{V},\\
\label{def_Gamma}
\Gamma &\equiv& V \frac{d^2 V/ d\phi^2}{(dV / d\phi)^2}.
\eeqn
Following Ref.~\cite{delaMacorra:1999ff},
we see that the equation for $H$  can also be written down as
\beqn
\label{b_eq_h}
H' = -3H[2X^2 + \gamma(1-X^2 -Y^2)].
\eeqn

Since the energy density of the barotropic fluid satisfies ($\rho_B \geq 0$), 
the contribution of the 
scalar field to the total energy density, 
$\Omega_{\phi} \equiv \rho_\phi/(6m_5^3 H) = X^2 + Y^2$
is bounded,$0 \leq X^2 + Y^2 \leq 1$. 
Hence the evolution of this system is completely described 
by trajectories within the unit disk.
Moreover, since the system is symmetric under the reflection
$(X, Y) \to (X, -Y)$ and the evolution will not go 
beyond the $Y=0$ line (which corresponds to 
$\phi = \infty$ or $H = \infty$), it is enough to discuss
the upper half disk $(Y \geq 0)$.

There is an intriguing aspect to these evolution equations. Given the definitions of $X$ and $Y$ above, Eqs.~(\ref{b_eq_x}-\ref{b_eq_l}) are identical in form to those for $X_{\rm CLW},\,Y_{\rm CLW}$ and $\lambda_{\rm CLW} \equiv -\frac{1}{\kappa_4}\frac{V'}{V}$ where $\lambda$ and $\lambda_{\rm CLW}$ are related through 
\beqn
\label{def_lclw}
\frac{\lambda}{\lambda_{\rm CLW}} &=& \frac{\kappa_4}{\kappa_5} \sqrt{\frac{2}{H}}\\ \nonumber
\eeqn 
in \cite{CLW,delaMacorra:1999ff,Ng:2001hs}. This means that the attractor solutions for $X_{\rm CLW}$ etc,  obtained in the standard Friedmann case where the Hubble parameter is determined by the linear energy density term, are the same attractor solutions in terms of our new variables $X$ etc... but for the case of $\rho^2$ domination. Even though they do not correspond to the same potential or evolution of the Hubble parameter, they do correspond to the same parameters in terms of $X,Y$ and $\lambda$. This is intriguing, as it implies there is a duality between the high energy $\rho^2$ set of solutions and the low energy solutions corresponding to $\rho$ domination.

\section{Attractor solutions}

We want to now turn our attention to the evolution equations (\ref{b_eq_x}) - (\ref{b_eq_l}). We are particularly interested in the case of scaling solutions, where $X'=Y'=\lambda'=0$.   To start with we consider the case where $\lambda$ tends to a constant value asymptotically. Using Eq.~(\ref{b_eq_y}), for non-trivial $X,Y$ and $\lambda$, we see that the condition $\lambda' =0$ in Eq.~(\ref{b_eq_l}) corresponds to the constraint $\Gamma = \frac{3}{2}$. Integrating, Eq.~(\ref{def_Gamma}) this in turn implies the form for the potential which leads to a constant $\lambda$, namely

\beqn
\label{pot_scaling}
V(\phi) = {\mu^6 \over  \phi^{2}}
\eeqn
where $\mu$ is a constant with the dimensions of mass. This result was first obtained using a different method in \cite{MMY}. The equivalent potential which leads to scaling solutions with constant $\lambda_{\rm CLW}$ in the case of the standard Friedmann cosmology is an exponential potential $V(\phi) = V_0 \exp(-\lambda_{\rm CLW} \kappa_4 \phi)$ \cite{CLW}, which is clear from Eqs.~(\ref{def_lambda}) and (\ref{def_lclw}). We can now summarise the attractor solutions $X_c,\,Y_c$, for our scaling potential Eq.~(\ref{pot_scaling}), recalling that they correspond to the same scaling solutions $X_{\rm CLW},\,Y_{\rm CLW}$ as found in \cite{CLW}. Eqs.~(\ref{b_eq_x}) and~(\ref{b_eq_y})
admit five different critical solutions for $X,~Y$ with $\lambda$ constant.

For these critical solutions, $(X_c=1,~Y_c=0)$,$(X_c=-1,~Y_c=0)$,
$(X_c=0,~Y_c=0)$ are unstable (extreme) critical points.
The other two solutions depend on the value of $\lambda$.
For $\lambda^2 > 3\gamma$,
we find the critical attractor values
\beqn
\label{scaling_sol}
X_c &=& \sqrt{\frac{3}{2}}\frac{\gamma}{\lambda},\nonumber\\
Y_c &=& \sqrt{\frac{3(2-\gamma) \gamma}{2\lambda^2}},
\eeqn
which satisfies $\gamma_\phi \equiv \frac{2X^2}{X^2 + Y^2}=\gamma$ implying that both the barotropic fluid and scalar field energy densities scale with the same redshift. 
The fractional energy density stored in the scalar field is given by 
$\Omega_{\phi c} = (3\gamma)/\lambda^2$.  If we 
consider the case of radiation for the barotropic fluid, then from Eqs.~(\ref{def_y}), (\ref{def_lambda}) and (\ref{scaling_sol}) we obtain 
$\Omega_{\phi c} = (1/2) (\mu/m_5)^3$ \cite{MMY}.
It is worth noting that because of the nature of the attractor solutions, 
for $3\gamma < \lambda^2
< 6 \gamma$, 
even if the scalar field dominates the universe,
the cosmic expansion law becomes the same as 
the barotropic fluid dominated universe.

If the asymptotic behavior of this system is characterized
by this scaling solution, then once the energy density has decreased to
$\rho \sim \rho_c$, we recover Einstein gravity with 
the linear energy density term dominating over  the quadratic term in Eqs.~(\ref{fr1}) and (\ref{fr2}). It is well known that in this conventional cosmology,
a scalar field with potential $V = \mu^6 \phi^{-2}$ leads to a
tracker solution and dominates the universe at late times 
\cite{Ratra_Peebles,Steinhardt}. In order for this term to then provide the observed Dark Energy today, it requires $\mu$ to be fine tuned to a value O(${\rm GeV}$), meaning that $\Omega_{\phi c}$ during the period that the quadratic term dominates  must be extremely small.

On the other hand, if $\lambda^2 < 6$,
the corresponding attractor solution is
\beqn
\label{p_l_sol}
X_c &=& \frac{\lambda}{\sqrt{6}},\nonumber\\
Y_c &=& \sqrt{1-\frac{\lambda^2}{6}},
\eeqn
which satisfies $\gamma_\phi = (1/3) \lambda^2$ and $\Omega_{\phi c} = 1$.
In this case the scalar field energy density dominates the universe 
asymptotically. 
If the scalar field has $\gamma_\phi < \gamma$,
then the solutions in Eqs.~(\ref{p_l_sol}) are stable, and the redshift
of the scalar field is slower than that of the barotropic fluid.
It corresponds to power-law inflation and 
is stable as in the conventional cosmology driven by
an exponential potential \cite{CLW,Lucchin}.
Even though this solution is interesting because it is exact,
it has the drawback of failing to stop inflating, so in a successful cosmology it would have to eventually  end inflation and recover the conventional cosmology. Furthermore, for such a large value of $\mu$, this scalar field dominates the universe much earlier.
However, if $\gamma_\phi= (1/3) \lambda^2 > \gamma$, 
then the solutions in Eqs.~(\ref{p_l_sol}) are unstable and 
the scalar field ends up in the regime of the solution
given in Eqs.~(\ref{scaling_sol}).

Having considered the case of a constant $\lambda$, we now turn our attention to the case where $\lambda(N)$ is not constant. Then the critical values in 
Eqs.~(\ref{scaling_sol}) and (\ref{p_l_sol}) hold
$X'~=Y'~=0$ only at single points, not over intervals of time.
This means that the attractor solutions to Eqs.~(\ref{b_eq_x})
and (\ref{b_eq_y}) are only valid as an asymptotic limit and
in general $X_c$, $Y_c$, $\lambda$ are time dependent.
If $X,~Y$ do not oscillate, then since their values are constrained to
$|X| \leq 1$, $|Y| \leq 1$, this implies that
they will approach  constant values asymptotically, 
given by the attractor solutions
of Eqs.~(\ref{b_eq_x}) and (\ref{b_eq_y}), and $X'$, $Y'$
will vanish. Therefore, we can generalize the attractor solutions 
of $X$, $Y$ given in Eqs.~(\ref{scaling_sol}) and (\ref{p_l_sol})
for more complicated potentials that have a nonconstant $\lambda(N)$.
In the following sections, we will analyze other asymptotic solutions
making use of this generalization.

\section{Potential dominated Inflationary solutions}

We now consider the case where  
$\lambda = - \frac{\sqrt{2}m_5^{3/2}}{\sqrt{H}}\frac{V'}{V}
\to 0$, asymptotically.
We might naively expect this happens as long as $V' \to 0$  faster than $V$ as $\phi \to \infty$, such as the case for inverse power law potentials of the form $V(\phi) = \mu^{4+n} \phi^{-n}$, with $n>0$.
Indeed, in the conventional Friedmann cosmology in which
$\lambda_{\rm CLW} \equiv -\frac{1}{\kappa_4}\frac{V'}{V}$
plays the same role as our $\lambda$, our expectation is realised in that  
$\lambda_{\rm CLW} \to 0$,
asymptotically \cite{delaMacorra:1999ff}.
However in our case,  if we substitute 
$V(\phi) = \mu^{4+n} \phi^{-n}$ with $n>0$
into Eq.~(\ref{def_lambda}),
and use the fact that in the regime where the potential is dominating, $H \propto V$,
we obtain $\lambda \propto \phi^{\frac{n}{2}-1}$.
From this, even with inverse power law potentials,
only when $2>n>0$, does $\lambda \to 0$ as 
$\phi \to \infty$.

In the following we will concentrate on models with potentials
which lead to  $\lambda \to 0$ asymptotically.
If this is realized, we can eliminate
the term proportional to $\lambda$
in Eq.~(\ref{b_eq_x}), and  Eq.~(\ref{b_eq_y}).
After rewriting them in terms of $H'$ with
Eq.~(\ref{b_eq_h}), and since $-6 < H'/H < 0$ for all values
of $X$, $Y$, and $\gamma$, we obtain the following
relations identical to those obtained in \cite{delaMacorra:1999ff}, but recalling the interpretation for the variables $X,\,Y$ and $\lambda$ differ in the two cases:
\beqn
\label{inf_cond}
\frac{X'}{X} &=& -\biggl(3 + \frac{1}{2} \frac{H'}{H}
\biggr) < 0,\nonumber\\
\frac{Y'}{Y} &=& -\frac{1}{2} \frac{H'}{H} > 0.
\eeqn

\begin{figure}[h]
\begin{center}
\includegraphics[width=80mm]{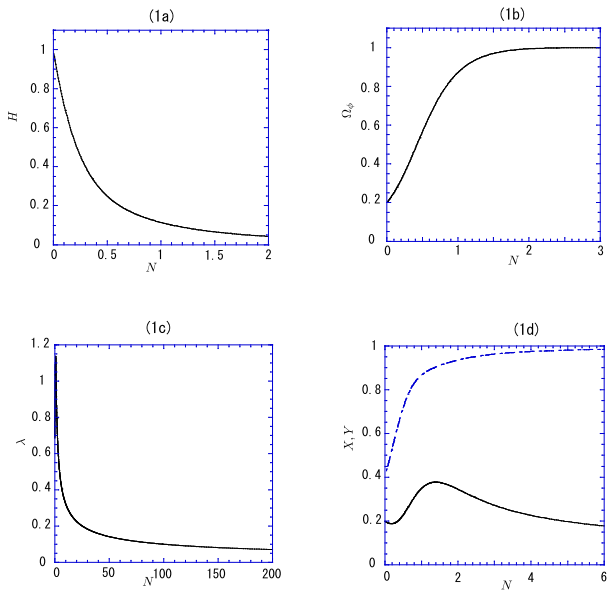}
\caption{Evolution of a Universe driven by the $\rho^2$ term in the Friedmann equation. The universe contains radiation
and a scalar field with $V = \mu^5 \phi^{-1}$. For simplicity, we choose 
$\mu = m_5$. The initial conditions 
are $X_0 = 0.2$, $Y_0 = 0.4$. 
In (a) 
the  Hubble parameter (solid curve) 
tends to a finite constant leading to inflation. 
In (b) we show 
$\Omega_{\phi}$ quickly approaching unity. In (c) $\lambda$
 evolves to zero asymptotically. In (d), the numerical
solutions for $Y$ (upper curve) and $X$ (lower curve) are plotted. The solutions
given by Eqs.~(\ref{inf_sol}) are obtained asymptotically.
}
\label{Figinflaiton}
\end{center}
\end{figure}

This of course is to be expected given the duality we discussed earlier. As explained in \cite{delaMacorra:1999ff} these equations indicate that $X \to 0$ its minimum value, as 
$Y \to 1$, its maximum value. This allows us to solve Eq.~(\ref{b_eq_x}), Eq.~(\ref{b_eq_y}), and 
Eq.~(\ref{b_eq_h}) for $X$, $Y$, and $H$ in the  region
$|X| \ll 1$, $|\lambda| \ll 1$ :
\beqn
\label{inf_sol}
X(N) &=& \frac{e^{-3N}}{\sqrt{1-c e^{-3\gamma N}}},
\nonumber\\
Y(N) &=& \frac{1}{\sqrt{1-ce^{-3\gamma N}}},
\nonumber\\
H(N) &=& d (1-e^{- 3 \gamma N}),
\eeqn
where $c$ and $d$ are integration constants.
These approximate solutions are consistent with those in Eq.~(\ref{p_l_sol}) in the limit $\lambda \to 0$, whilst recalling that $X$ and $Y$ are varying with time. They show that 
the scalar field potential dominates
the energy density of the universe as the universe inflates with 
almost constant  Hubble parameter. 
In fig.~\ref{Figinflaiton},  we demonstrate the validity of the solutions in Eq.~(\ref{inf_sol}), as we show numerical solutions for a model with a barotropic fluid of radiation ($\gamma =4/3$), and with potential 
$V(\phi) = \mu^5\phi^{-1}$. 

This asymptotically inflating  solution is stable, hence in order 
to recover standard cosmology at late times, a mechanism
to end this period of inflation is necessary. Simply relying on the existence of the $\rho$ term dominating in the Friedmann equation will not be sufficient here, because as shown in \cite{delaMacorra:1999ff}, this potential also has an inflationary solution in that era. 
This  can be readily seen from Eq.~(\ref{def_lclw}), where 
$\lambda_{\rm CLW} = (\kappa_5/\kappa_4) \sqrt{H/2} \,\lambda \to 0$
as $\lambda \to 0$ for constant $H$.

\section{Kinetic-term dominated solution}

Having considered the case $\lambda \to$ const (including zero), we now consider the case where asymptotically $\lambda = - \frac{\sqrt{2}m_5^{3/2}}{\sqrt{H}}\frac{V'}{V}  \to \infty$ smoothly without oscillations. 

The presence of the $1/\sqrt{H}$ term in the definition of $\lambda$, as opposed to the definition of $\lambda_{\rm CLW} \propto V'/V$, and the fact that in most cosmologies $H \to 0$ asymptotically, means that a wider class of potentials satisfy $\lambda \to \infty$, than satisfy  $\lambda_{\rm CLW} \to \infty$ \cite{delaMacorra:1999ff}. For example, as we show below, the inverse polynomial potentials, $V(\phi) = \mu^{4+n} \phi^{-n}$ with $n>2$ discussed earlier 
satisfy this condition. Another concrete example is 
a model including an exponential potential which satisfies 
$\lambda \propto 1/\sqrt{H} \to \infty$.

Although the specific potentials included in this limit may be different between the $\rho^2$ and $\rho$ cosmologies,  the invariance of the form of the equations of motion (\ref{b_eq_x})- (\ref{b_eq_y}) when written in terms of $X$ and $Y$ implies that the solutions obtained in terms of $X$ and $Y$ in Ref.~\cite{delaMacorra:1999ff} for the case of the conventional $\rho$ dominated cosmology, apply to our case aswell. We therefore summarise the results presented there. 

\begin{figure}[h]
\begin{center}
\includegraphics[width=80mm]{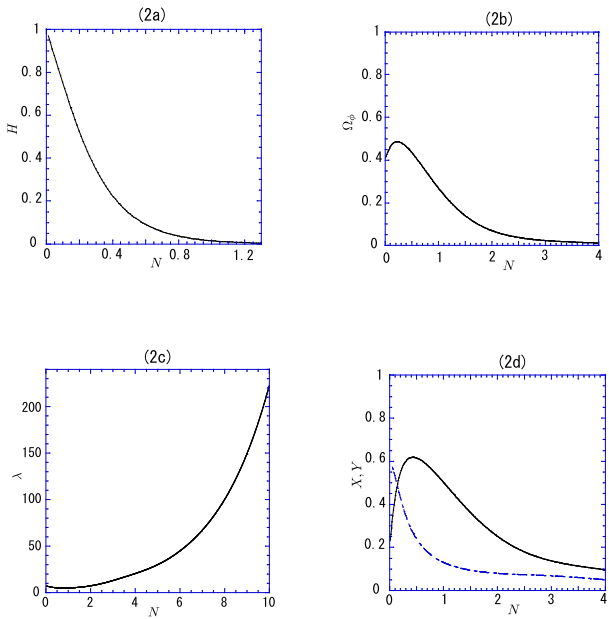}
\caption{Evolution of the $\rho^2$ cosmology for a universe filled with radiation
and a scalar field with $V = \mu^7 \phi^{-3}$. For simplicity, we choose $\mu = m_5$.
The initial conditions 
are $X_0 = 0.2$, $Y_0 = 0.6$. In (a) we show
that the  Hubble parameter normalized by its initial value (solid curve) 
tends to zero. In (b) we show 
$\Omega_{\phi}$ quickly approaching $0$. In (c) $\lambda$
evolves to infinity asymptotically. In (d), the numerical
$X$ (solid curve) and $Y$ (dotted curve) solutions are plotted. 
As in Eqs.~(\ref{kin_ap}), at the beginning, $X$ increases and
reaches its maximum value, before turning over following 
Eq.~(\ref{kin_mid}) as it (and $Y$) approaches $0$ asymptotically.
}
\label{Figkinetic}
\end{center}
\end{figure}

Initially we expect $X$ and $Y$ to be order unity, which means for  $|\lambda| \gg 1$, 
the leading terms of Eq.~(\ref{b_eq_x}) 
and Eq.~(\ref{b_eq_y}) are
\beqn
\label{kin_ap}
X' &=& \sqrt{\frac{3}{2}} \lambda Y^2, \nonumber\\
Y' &=& -\sqrt{\frac{3}{2}} \lambda XY. 
\eeqn
Note for positive (negative) $\lambda$,  $X \to 1 (-1)$ whereas for both signs, 
$Y \to 0$. Eventually as $Y$ keeps decreasing, other terms in Eqs.~(\ref{b_eq_x}) and  (\ref{b_eq_y}) become important. In particular if $|X'| \geq |\lambda| Y^2$,
then using Eq.~(\ref{b_eq_h}), the evolution of $X$
is given by 
\beqn
\label{kin_mid}
\frac{X'}{X}  = - \biggl(3+\frac{H'}{2H}\biggr). 
\eeqn
Now since  $-6 < H'/H < 0$, $X$ having reached a maximum value turns over and like $Y$, heads off finally approaching
the values given by the solution of
Eqs.~(\ref{scaling_sol}), $X \to X_c$, $Y \to Y_c$.
For $X_c$ and $Y_c$ with $|\lambda| \gg 1$, then 
$1 \gg |X_c| > Y_c$  holds, and $\rho_\phi$ will decrease faster than $\rho_B$ with $\Omega_\phi \to 0$.
Even though $X_c$, $Y_c$ are not critical 
(constant) points since $\lambda$ is not constant,
we have verified numerically that  the above asymptotic solutions are good approximations. In Fig.~\ref{Figkinetic}, typical examples of 
the numerical results for a model with potential 
$V(\phi) = \mu^7\phi^{-3}$ are shown, in a universe containing radiation.  Note in particular the behaviour of $\lambda \to \infty$ in (c) and how in (d) $X$ increases initially before decreasing towards zero, approaching the solutions given in Eqs.~(\ref{scaling_sol}). As expected these solutions match closely those presented in Figure 2 of  Ref~\cite{delaMacorra:1999ff}, for the conventional $\rho$ dominated cosmology, but there are important differences. In particular the form of the potential involved (exponential in their case, inverse polynomial in ours), and most significantly the fact we are dealing with different cosmologies. 

Let us briefly digress to discuss the impact of this sort of solution. We have seen that as long as $\mu \leq m_5$, then the asymptotic behaviour in the $\rho^2$ cosmology, is one where $\rho_{\gamma} >\rho_{\phi}$. This will carry on until $\rho \sim \rho_c$, after which we will enter a conventional radiation dominated universe. This is a nice feature as once in that regime we know from earlier studies that the inverse polynomial potentials being discussed here lead to tracking behaviour necessary to explain the dark energy today. The power of the $\rho^2$ regime here lies in the fact that it provides a dynamical explanation of the large difference between the 
energy density of the dominant barotropic fluid 
and the scalar field `initially' as we enter the conventional
cosmology \cite{Maeda:2000mf}.

If we decide to concentrate initially on the limit $\phi \to 0$,
then this corresponds to  $\lambda \to 0$ 
suggesting an unstable inflationary solution in this limit,
for the inverse polynomial potentials  
$V(\phi) = \mu^{4+n} \phi^{-n}$, with $n>2$.
This result provides one of a number of possibilities which have been investigated where the same scalar field potential is used to realize both a period of early inflation
and late time dark energy domination, although there are strong constraints on their viability arising from the tendency to overproduce gravitational waves during reheating.
This is a form of quintessential inflation \cite{Peebles:1998qn}, except that it is making use of the brane-world scenario, and has been investigated recently by a number of authors  
\cite{Copeland:2000hn,Sahni:2001qp,Majumdar:2001mm,Nunes:2002wz,Lidsey:2003sj,Tsujikawa:2003zd,Sami:2004ic,MMY}.

\section{Oscillating solutions}

We now turn our attention to consider the case $|\lambda| \to \infty$
with the scalar field $\phi$ oscillating about the minimum of its  potential asymptotically. In the context of realistic scenarios this is somewhat of a formal exercise in that we might well expect that the $\rho \gg \rho_c$ constraint would have been violated once this situation had been reached. However, we feel that it is worth investigating in its own right as it allows us to pursue the duality relation we have identified in the paper between the two regimes of $\rho$ domination and $\rho^2$ domination. 

Given that $\phi$ is oscillating about its minimum, we can  
without loss of generality, take
it to be zero, and expand $V(\phi)$ as a power series about the minimum.
Keeping only the leading term, 
we have $(1/n) \mu^{4-n} \phi^n$,
 where $n = 2, 4, 6, \cdots$, even because of the boundedness of the potential.  
(For $n=4$ we use 
$V(\phi) = (1/4) \hat{\lambda} \phi^4$ where $\hat{\lambda}$ is 
a dimensionless constant. )

Again because of the duality invariance, the equations match those in \cite{delaMacorra:1999ff}, so we follow their path to determine under which conditions, $\Omega_\phi$:
either dominates (goes to unity); oscillates around a finite constant value;
or vanishes asymptotically.
In actual fact in \cite{MMY}, we have already analyzed the cases for $n=2$ and $4$ 
obtaining analytic solutions by making assumptions about which components of the energy density dominate the universe, initially.
In these cases, we have obtained approximate solutions
by invoking the virial theorem, in which the relation between 
the time-averaged value of the kinetic term and
the potential term of the energy density of the scalar field is given as
\beqn
\label{virial}
\langle \dot{\phi}^2 \rangle = n \langle V(\phi) \rangle.
\eeqn

From Eq.~(\ref{virial}), it is clear that the scalar field behaves as a pressureless
perfect fluid (dust fluid) ($\langle \gamma_\phi \rangle = 1$) 
for $n = 2$,  while it behaves as a radiation fluid 
($\langle \gamma_\phi \rangle = 4/3$) for $n=4$.
Therefore, if we consider the case of radiation for
the barotropic fluid in the early universe, 
for the case $n = 2$, then the energy density in the scalar field will decrease slower than that of radiation, leading to  $\Omega_\phi \to 1$. For the $n = 4$, case, both fluids evolve at the same rate, hence $\Omega_\phi \to cte$, a constant value
determined  by the self coupling parameter $\hat{\lambda}$ in $V(\phi)$.

In order to analyze the case for general $n$, 
and without specifying which component dominates the universe,
we will introduce the total adiabatic index $\gamma_{\rm tot}$.
When it is the barotropic fluid that dominates, then 
$\gamma_{\rm tot} = \gamma$, but when the scalar field
dominates the universe, we take 
$\gamma_{\rm tot} = \langle \gamma_\phi \rangle$ since in this case
$\gamma_{\phi}$ oscillates. 
Now, in terms of $\gamma_{\rm tot}$, $H(t) = 1/(3 \gamma_{\rm tot} t)$,
hence in the $\rho^2$ dominated era, Eq.~(\ref{KK}) can be rewritten
as,

\beqn
\label{k-gordon}
\ddot{\phi} + \frac{1}{\gamma_{\rm tot} t}\dot{\phi} +  \mu^{4-n} \phi^{n-1}=0.
\eeqn
(Recall for the case $n=4$, the third term must be $\hat{\lambda} \phi^3$.
In that case, if we introduce a new time variable 
$\tau \equiv \sqrt{\hat{\lambda}} t$, the differential equation 
takes the same form as Eq.~(\ref{k-gordon}) with $n=4$. Where appropriate 
we use this time variable for the case $n=4$.)

Even though we have already obtained the results for the case $n=2$
in \cite{MMY}, we first consider this case to confirm our previous results. 
For $V(\phi) = (1/2) \mu^2 \phi^2$,
the solution of Eq.~(\ref{k-gordon}) is given in terms of
$J_m$ and $Y_m$, the Bessel functions of the first and second kind,
respectively, 
\beqn
\phi(z) &=& \mu z^{m} 
\left[ c_1 J_m(z) + c_2 Y_m(z)\right],\nonumber\\
\dot{\phi} (z) &=& \mu^2 z^{m}
\left[ c_1 J_{m-1}(z) + c_2 Y_{m-1}(z)\right],
\label{bessel}
\eeqn
where $c_1$, and $c_2$ are constants,
$m \equiv (1/2)-(1/2\gamma_{\rm tot})$,
$z \equiv \mu t$
\cite{Liddle}.
From Eq.~(\ref{bessel}), $X$ and $Y$ are expressed as 
\beqn
\label{bessel_xy}
X &=& r z^{m+\frac{1}{2}} [k J_{m-1}(z) + Y_{m-1}(z)],\nonumber\\
Y &=& r z^{m+\frac{1}{2}} [k J_m(z) + Y_m(z)],
\eeqn
where $r \equiv (1/2) (\mu/m_5)^{3/2} \sqrt{\gamma_{\rm tot}} c_2$,
$k \equiv c_1/c_2$. A simple analytic expression can be obtained using 
the formula for the asymptotic limit of the Bessel functions
$J_m \sim \sqrt{(2/\pi z)} \cos [z-\pi (2m+1)/4]$,
$Y_m \sim \sqrt{(2/\pi z)} \sin [z-\pi (2m+1)/4]$ 
for $z \gg 1$. The amplitude of $X$ and $Y$
in Eqs.~(\ref{bessel_xy}) in the limit $z \gg 1$
goes as $X \sim Y \sim z^{m}$. A finite value of $X$, $Y$ requires
$m = (1/2)-(1/2\gamma_{\rm tot}) = 0$ (i.e. $\gamma_{\rm tot} =1$).
Next, we analyze how the asymptotic value of $\gamma_{\rm tot}$ are 
decided. In this case, regardless of the form of the barotropic fluid,
in the limit $z \gg 1$, 
$\langle X^2 \rangle$ $=$ $\langle Y^2 \rangle$ $=$
$(r^2/\pi) z^{2m} (k^2 +1)$, which yields 
$\langle \gamma_\phi \rangle = 1$. 
This is completely consistent with the previous results
relying on the virial theorem. We can confirm, therefore,
if the barotropic fluid has $\gamma < 1$, i.e.
smaller than $\gamma_\phi$, then the barotropic fluid decays
slower than the scalar field, which yields 
$\gamma_{\rm tot} < 1$. In this case, from the asymptotic limit
of the Bessel functions, $\Omega_\phi = X^2 + Y^2 \to 0$.
On the other hand, if the barotropic fluid has 
$\gamma > 1$, i.e. greater than $\gamma_\phi$, then  
the barotropic fluid decays faster than the scalar field.
In this case, 
$\gamma_{\rm tot} \to \langle \gamma_\phi \rangle = 1$, which leads to 
$\Omega_\phi = X^2 + Y^2 \to 1$ asymptotically.
For the case $\gamma = 1=\gamma_{\rm tot} $,  
$\Omega_\phi$ asymptotes to some constant value
between $0$ and $1$.

We have obtained an analytic solution of Eq.~(\ref{k-gordon})
for the case $V(\phi) = (1/2) \mu^2 \phi^2$. 
This is clearly the simplest case since
Eq.~(\ref{k-gordon}) is linear in $\phi$ and its derivatives.
For $V(\phi) = (1/n) \mu^{4-n} \phi^{n}$ with $n>2$,
Eq.~(\ref{k-gordon}) becomes nonlinear in $\phi$
and no simple analytic solution exists. 
However, using the following ansatz,
\beqn
\label{osc_phi}
\phi = \mu z^{-\frac{1}{n}} \left[c_1 \cos \left(\beta 
z^{\frac{(n+2)}{2n}}\right)
+ c_2 \sin \left(\beta z^{\frac{(n+2)}{2n}}\right)\right],
\eeqn
we obtain a solution with the correct asymptotic behavior $\phi^{n/2}$, 
$\dot{\phi} \propto t^{-1/2}$, $H \propto t^{-1}$,
and $\ddot{\phi} \propto \phi^{n-1}$, necessary if we wish $X$, $Y$
to have finite values other than $0$ and $1$. 
The $z (=\mu t)$ exponents in Eq.~(\ref{osc_phi}) are such that  
Eq.~(\ref{k-gordon}) is satisfied up to leading orders $t^{-3/2}$ and 
$t^{(1-n)/n}$. The same condition also imposes the conditions 
$\gamma_{\rm tot} = 2n/(2+n)$ and 
$\beta^2 =  (2n/n+2)^2 c_2^{n-2} 
[\sin (\beta z^{2/n}) + k \cos (\beta z^{2/n})]^{n-2}$.
Notice that the value of $\gamma_{\rm tot}$ is precisely the one  
we obtained earlier based on general arguments 
[cf. Eq.~(\ref{virial})]. Of course, the ansatz in 
Eq.~(\ref{osc_phi}) is not a complete solution to Eq.~(\ref{k-gordon}) as can be seen in the fact that  $\beta$ is not a true constant (except for the case $n=2$). For $n > 2$ we must use the average value of $\beta$ evaluated in the asymptotic region. 
In terms of Eq.~(\ref{osc_phi}), $X$ and $Y$
take the following expressions:
\beqn
\label{osc_xy}
X &=& X_0 \left[\cos \left(\beta 
z^{\frac{(n+2)}{2n}}\right)
- k \sin \left(\beta z^{\frac{(n+2)}{2n}}\right)\right],
\nonumber\\
Y &=& Y_0 \left[\sin \left(\beta 
z^{\frac{(n+2)}{2n}}\right)
+ k \cos \left(\beta z^{\frac{(n+2)}{2n}}\right)\right]^{\frac{n}{2}},
\eeqn
with $X_0 =\frac{(n+2)}{4n} \sqrt{\gamma_{\rm tot}} \beta (\mu/m_5)^{3/2} c_2$
and $Y_0 = \sqrt{\gamma_{\rm tot}/2n} (\mu/m_5)^{3/2}c_2^{n/2}$.

From Eq.~(\ref{osc_xy}) we obtain 
$\langle Y^2 \rangle / \langle X^2 \rangle = 2/n$,
asymptotically, as in Eq.~(\ref{virial}),
which is consistent with the virial theorem.
We have, therefore, $\langle \gamma_\phi \rangle
= 2n/(2+n)$ and $\gamma_{\rm tot} = \langle \gamma_\phi \rangle$,
i.e., the ansatz in Eq.~(\ref{osc_phi}) is a solution to 
Eq.~(\ref{k-gordon}) only when the dominant energy density redshifts
as fast as the scalar field. This is, of course, no surprise 
since we imposed on the ansatz Eq.~(\ref{osc_phi}),
the limit $X$, $Y$ $\to cte$ ($\neq 0$ or $1$).

\begin{figure}[h]
\begin{center}
\includegraphics[width=80mm]{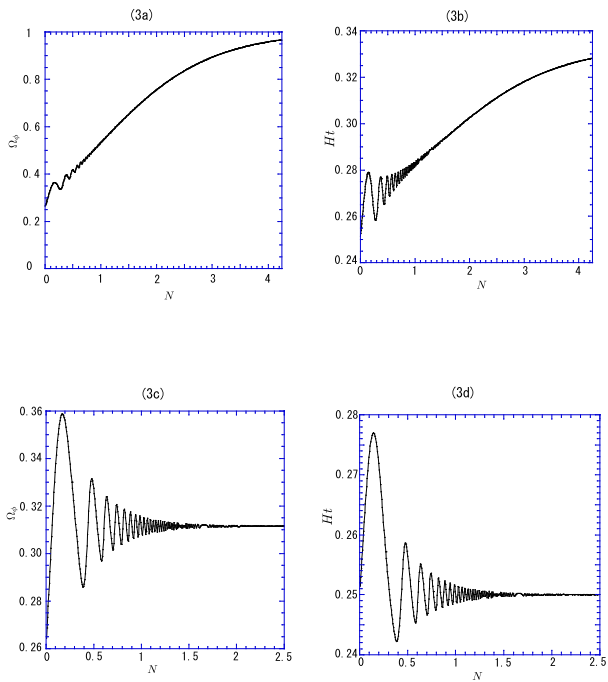}
\caption{
Evolution of the universe filled with radiation
and a scalar field with potential $V = (1/2)\mu^2\phi^2$ ((a), (b))
and $V = (1/4)\hat{\lambda} \phi^4$ ((c), (d))
in the $\rho^2$  dominating  era. 
For simplicity, we choose $\mu = m_5$ and $\hat{\lambda} = 1$.
The initial conditions are $X_0 = 0.1$, $Y_0 = 0.5$. 
We confirm, as expected, that the asymptotic behavior of the oscillating
scalar field is characterized by $\langle \gamma_{\phi} \rangle = 2n/(2+n)$.
In (a) we show the time evolution of $\Omega_\phi$ with a $\phi^2$ potential ($n=2$).
Since the oscillating scalar field behaves like dust 
for $n=2$, it gradually dominates the radiation and 
$\Omega_\phi$ oscillates, asymptoting to unity.
Similar behavior is seen in $Ht$ in (b), where it approaches
to $1/3$ corresponding to the cosmic expansion law
of a dust dominated universe in the $\rho^2$ dominating era.
In (c) we show the time evolution of $\Omega_\phi$ for the $\phi^4$ potential, ($n=4$). Since the oscillating field behaves 
like radiation here, after a few oscillations, it approaches a constant value. In (d), for $n=4$, $Ht \to \frac{1}{4}$, 
corresponding to the cosmic expansion law
of a dust dominated universe in the  $\rho^2$ dominated era.
}
\label{oscillation}
\end{center}
\end{figure}

Even though the above discussion strictly relies on the time-averaged value,
we have confirmed that it provides a good approximation to the true numerical results.
In fig.~\ref{oscillation}, typical examples of the numerical results
for models with $V(\phi) = (1/2) \mu^2\phi^2$ (in (a) and (b))
and $V(\phi) = (1/4) \lambda \phi^4$ (in (c) and (d))
are plotted. As for the barotropic fluid, we use radiation 
($\gamma = 4/3$). For both potentials, we draw the time evolution of 
$\Omega_\phi$ and $Ht$.

To conclude this part of the analysis, 
if the initially dominant energy density component 
has a larger (smaller) adiabatic index than
$\langle \gamma_\phi \rangle = 2n/(2+n)$, 
then $\Omega_{\phi}$ will approach $1$ ($0$).
For example,  if we consider radiation as
the barotropic fluid, then for $n > 4$, the energy density 
of the scalar field will decrease faster than radiation,
which guarantees the recovery of the standard cosmology
at an epoch when  the energy density is lower than $\rho_c$.

Even though the form of these relations are the same as in 
a conventional cosmology, the oscillating behavior
found in the regime dominated by the quadratic term has a distinct  advantage
in the preheating phase \cite{b_pre} if this scalar field couples 
to other matter fields. Since the cosmic expansion at late times 
becomes slower and the amplitude of $\phi$ decreases more slowly than in 
the conventional case, there may well be  sufficient particle production generated as the field decays.

\section{Summary}

In this paper, we have studied the novel dynamics of a scalar field plus barotropic fluid, 
in a brane-world scenario dominated by the $\rho^2$ term in the effective four-dimensional Friedmann constraint. 
As a concrete example,
we have adopted the Randall-Sundrum II model and
assumed that the scalar field is confined to our four dimensional
spacetime. 

Our approach has allowed us to deal with general classes of potentials, and complements an earlier investigation of a similar system but for particular potentials in \cite{MMY}. 
Perhaps the most important result we have obtained, can be seen in the defining equations (\ref{def_x})-(\ref{b_eq_h}) in which we introduce a new set of variables to analyse the evolution equations in a model independent manner. The crucial point is that the equations for $X$ and $Y$, given by  (\ref{b_eq_x}) and (\ref{b_eq_y}) are identical to those derived in the conventional four dimensional cosmology where the Friedmann equation is driven by the energy density $\rho$ of the scalar field and barotropic fluid, \cite{CLW,delaMacorra:1999ff,Ng:2001hs} as opposed to the $\rho^2$ driving term in our case. Although the precise definitions of $X$ and $Y$ differ, the fact that they obey the same evolution equation allows us to immediately write down and understand the form of the scaling solutions. There is a simple map which allows us to relate the cosmological solutions in each regime -- a duality between the parameters $X,\,Y$ in the two cases. 
\begin{table}[t]
\begin{tabular}{cccc}
\hline
\hline
$\lambda = - \frac{\sqrt{2}m_5^{3/2}}{\sqrt{H}}\frac{V'}{V}$     & $\Omega_{\phi} = \frac{\rho_{\phi}}{\rho}$       & $\gamma_{\phi}$ & $V(\phi)$ \\
\hline
\\
$c={\rm cte}\;\;(> \sqrt{\frac{3\gamma}{2}})$     & $\frac{3\gamma}{2c^2}$      &  $\gamma$  & $\mu^6 \phi^{-2}$\\
$c={\rm cte}\;\;(< \sqrt{\frac{3\gamma}{2}})$& $1$ &  $\frac{2c^2}{3}$ & $\mu^6 \phi^{-2}$ \\
$\infty\;\;({\rm no\,oscil.})$          & $0 $         &  $ \gamma $  &  $\mu^{n+4} \phi^{-n},\;\;n>2$    \\
$\infty\;\;({\rm oscil.})$           & $0$     &  $\frac{2n}{2+n}\;\;(>\gamma)$  &  $\frac{\mu^{4-n}}{n} \phi^n,\;n>0,\;{\rm even}$ \\
~         & ${\rm cte}$    &  $\frac{2n}{2+n}\;\;(=\gamma)$ & ~    \\
~ & 1 & $\frac{2n}{2+n}\;\;(<\gamma)$ & ~ \\
0 & 1& 0 & $\mu^{n+4} \phi^{-n},\;\;2>n>0$\\
\\
\hline
\hline
\end{tabular}
\caption{\label{table1}The asymptotic behavior 
of the contribution of the scalar field to the total energy density
$(\Omega_{\phi})$ and the effective equation of state for 
the scalar field $(\gamma_{\phi})$ for different limiting cases of
$\lambda(\phi)$ in a $\rho^2$ dominated cosmology. In the last column we give an example of
the potential $V(\phi)$ which satisfies this limit. Note how the duality transformation implies solutions match those in \cite{delaMacorra:1999ff} for the variables $\lambda, \,X$ and $Y$, although the definitions of those variables differ. In particular, the corresponding potentials in the last column differ for the two cases.}
\end{table}

We have  summarized our results in Table I, in an analogous manner to that presented in \cite{delaMacorra:1999ff} for the equivalent conventional $\rho$ dominated cosmology. In particular 
we showed that all the model dependence is given by 
$\lambda = - \frac{\sqrt{2}m_5^{3/2}}{\sqrt{H}}\frac{V'}{V} $ and 
the adiabatic index of the barotropic fluid $\gamma$.
Scalar potentials that do not require introducing nonzero minima
are classified into one of three different limiting cases
by the asymptotic behavior of $\lambda$:
$\lambda$ goes to a finite constant, zero, or infinity.
In the first case, $\Omega_{\phi}$ approaches a finite constant 
(different from one or zero) depending on the value of $\lambda$.
It is worth noting that in the conventional cosmology this happens 
in the model with an exponential potential, 
$V(\phi) = \mu^4 \exp~(-\lambda_{\rm CLW} \kappa_4 \phi)$, while in the brane-world 
cosmology, it happens with an inverse square potential model,
$V(\phi) = \mu^6 \phi^{-2}$ -- a reflection of the duality relating $X,\,Y$ and $X_{\rm CLW},\,Y_{\rm CLW}$.

In the second case $\lambda \to 0$,
we obtained $X \to 0$, $Y \to 1$ 
corresponding to inflation with an almost constant Hubble parameter.
As a concrete example, 
we considered the model with $V(\phi)= \mu^{4+n} \phi^{-n}$. 
For this potential, in the conventional cosmology, 
the inflationary solution is an attractor for any $n>0$,
but in the presence of the quadratic density term, it occurs only for 
$2>n>0$. 

In the final case $\lambda \to \infty$. Following the procedure in \cite{delaMacorra:1999ff} we investigated two different
possibilities depending on whether $\lambda$ oscillated or not.
If it didn't, $X$, $Y$, $\Omega_{\phi} \to 0$
asymptotically. We showed the model with an exponential potential
belongs to this class in the presence of the quadratic 
energy density term, as do models with inverse power law potentials with $n>2$.
As pointed out in \cite{Maeda:2000mf}, this 
provides a new feature for the quintessence scenario in that it allows us to explain how the energy densities in the radiation and scalar field could be different as we enter the usual $\rho$ dominated Friedmann cosmology era.  
On the other hand, if $\lambda$ does oscillate, the adiabatic index
of the scalar field is given as 
$\langle \gamma_\phi \rangle = 2n/(2+n)$,
where $n$ is the power of the leading term in the scalar potential.
The asymptotic behavior of the universe is decided by whether 
$\langle \gamma_\phi \rangle$ is larger than $\gamma$ or not.
For $\langle \gamma_\phi \rangle > \gamma$, $\Omega_{\phi} \to 0$, while for $\langle \gamma_\phi \rangle < \gamma$, $\Omega_{\phi}\to 1$ and for $\langle \gamma_\phi \rangle \gamma$, $\Omega_{\phi}\to cte$. 

Although, we have concentrated on investigating the cosmology in the $\rho^2$ dominated regime, any realistic cosmology also has to allow for the fact that the universe is dominated today by the standard $\rho$ term in the Friedmann equation. The next thing to do is to match the two regimes together. This has been investigated by a number of authors \cite{Copeland:2000hn,Sahni:2001qp,Majumdar:2001mm,Nunes:2002wz,Lidsey:2003sj,Tsujikawa:2003zd,Sami:2004ic,MMY} mainly using numerical simulations involving particular potentials. Such quintessential inflation models are being tightly constrained by recent CMBR data as they tend to generate too large a tensor contribution to the CMBR power spectra \cite{Sahni:2001qp,Sami:2004ic}. What we have developed in this paper is an alternative, possibly powerful approach which allows us to make use of the duality that exists between the two regimes ($\rho^2$ and $\rho$ dominated). We are currently investigating using this duality to determine consistent cosmologies involving an evolution from one regime into the other in a model independent manner, but based on the idea of using the definition of $\lambda$ and $\gamma$ as the key ingredients in the analysis.

\section*{Acknowledgments}

S.~M.~would like to thank Kei-ichi Maeda for the continuous
encouragement. He is grateful to the University of Sussex for their hospitality during a period when this work was initiated. S.~J.~L.~would like to acknowledge the support of the Overseas Research Students Awards for financial support. This work was partially supported by 
 The 21st century COE Program (Holistic Research and Education Centre
for Physics self-organization systems at Waseda University.)


\end{document}